%% file: mm036-althoff.tex
\documentclass{sig-alternate-2013}

\usepackage[T1]{fontenc}

\newfont{\mycrnotice}{ptmr8t at 7pt}
\newfont{\myconfname}{ptmri8t at 7pt}
\let\crnotice\mycrnotice%
\let\confname\myconfname%

\permission{Authors' pre-submission draft.}
\conferenceinfo{}{{\mycrnotice{}}}
\copyrightetc{}
\crdata{}

\usepackage{csvsimple}
\usepackage{subfig}
\usepackage{hyperref}
\usepackage{multirow}
\usepackage{color}

\usepackage{array}

\DeclareMathOperator*{\median}{median}
\DeclareCaptionType{copyrightbox}

\def\sharedaffiliation{%
\end{tabular}
\begin{tabular}{c}}

\begin{document}
%

\title{Analysis and Forecasting of Trending\\ Topics in Online Media Streams}

\numberofauthors{1}
\author{
\alignauthor
Tim Althoff \quad\quad
Damian Borth \quad\quad
Jörn Hees \quad\quad
Andreas Dengel
\sharedaffiliation
	\affaddr{German Research Center for Artificial Intelligence (DFKI)}\\
	\affaddr{D-67663 Kaiserslautern, Germany}\\
	\email{althoff@stanford.edu, \{damian.borth, joern.hees, andreas.dengel\}@dfki.de}
}

\maketitle
\begin{abstract}
Among the vast information available on the web, social media streams capture what people currently pay attention to and how they feel about certain topics. Awareness of such trending topics plays a crucial role in multimedia systems such as trend aware recommendation and automatic vocabulary selection for video concept detection systems.

Correctly utilizing trending topics requires a better understanding of their various characteristics in different social media streams. To this end, we present the first comprehensive study across three major online and social media streams, Twitter, Google, and Wikipedia, covering thousands of trending topics during an observation period of an entire year. Our results indicate that depending on one's requirements one does not necessarily have to turn to Twitter for information about current events and that some media streams strongly emphasize content of specific categories. As our second key contribution, we further present a novel approach for the challenging task of forecasting the life cycle of trending topics in the very moment they emerge. Our fully automated approach is based on a nearest neighbor forecasting technique exploiting our assumption that semantically similar topics exhibit similar behavior.

We demonstrate on a large-scale dataset of Wikipedia page view statistics that forecasts by the proposed approach are about 9-48k views closer to the actual viewing statistics compared to baseline methods and achieve a mean average percentage error of 45-19\,\% for time periods of up to 14 days.
\\
\end{abstract}




\input{introduction}
\input{relatedwork}
\input{analysis}

\input{forecasting}

\input{evaluation}

\input{conclusion}


%

\small
\bibliographystyle{abbrv}

%
%
\end{document}

%% file: introduction.tex
\section{Introduction}

\begin{figure}
\centering
\includegraphics[width=\columnwidth]{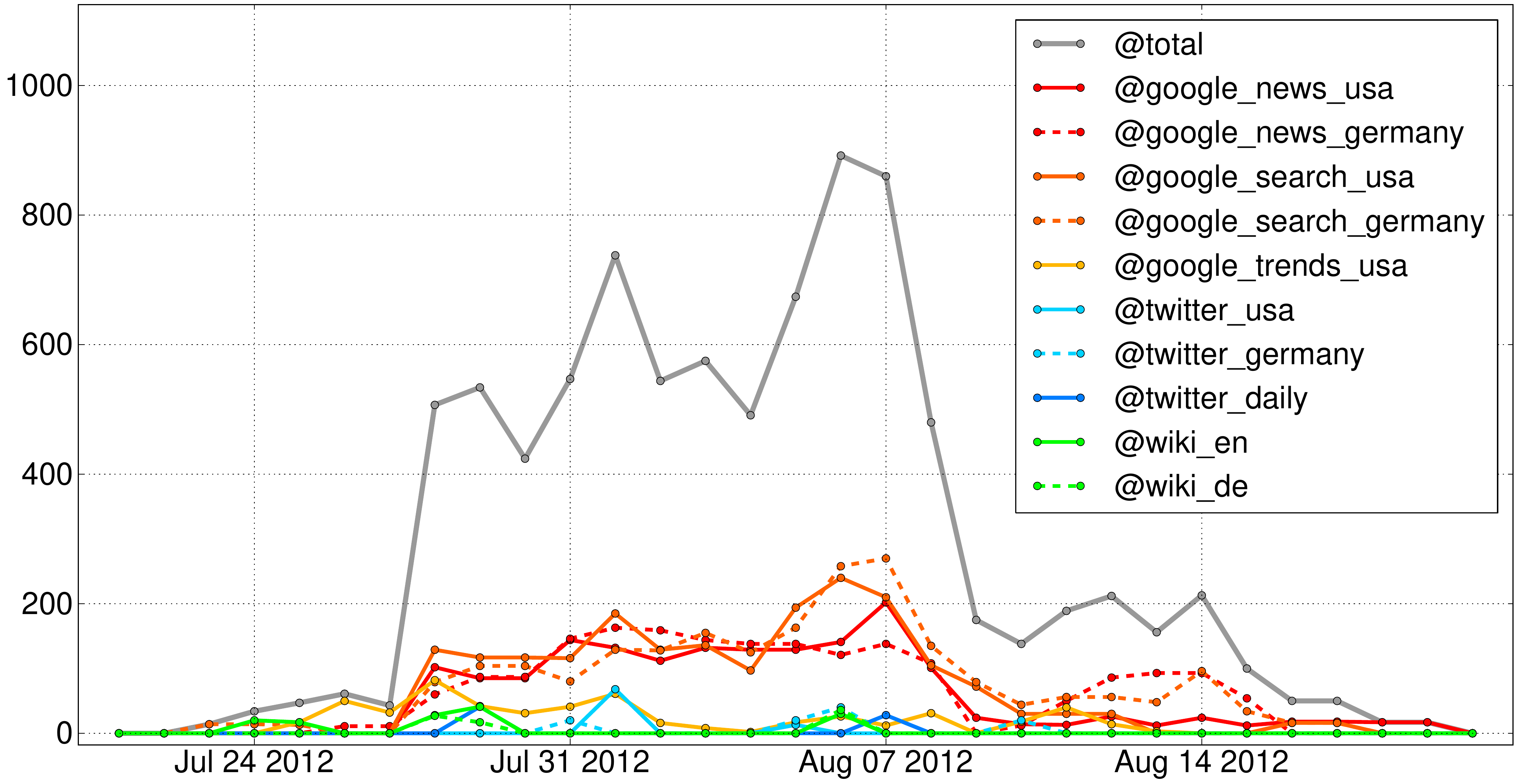}
\caption{The trending topic ``Olympics 2012'' during summer 2012. The colored curves represent the contribution of the different online and social media streams.}\label{fig:olympics}
\end{figure}

Currently, as multimedia content consumption is rapidly increasing on the internet we can consider ourselves as living in the zettabyte era\footnote{Cisco, The Zettabyte Era. May 30, 2012.}.
To a large degree this vast amount of information is created, shared, and exchanged by the users themselves in social multimedia systems. These systems provide us with streams of information that capture how we are spending our time and what we are talking about. Over time, different topics arise in these media streams reflecting changing interests of groups of individuals. In this sense, social media streams mirror our society. These streams contain rich information and immediate feedback about to what people currently pay attention to and how they feel about certain topics. 

Such \textit{trending topics} \cite{naaman2011hip} can be associated with a subject (i.e. a textual label) and experience sudden increases in user popularity (``trending''). They usually correspond to real world events such as sport events (Olympics 2012), product releases (iPhone 5), celebrity news (Steve Jobs' death), incidents (Sinking of the Costa Concordia), political movements (Occupy), and entertainment (Academy Awards). 
Awareness of trending topics plays a key role in building social multimedia systems satisfying users' information needs and enhancing the multimedia consumption experience \cite{roy2012socialtransfer}. 

Today, a large variety of social multimedia systems is available to users on the web such as Twitter, YouTube, Facebook, Flickr, Pinterest, Tumblr, Google News/Trends, and Wikipedia serving different kinds of needs such as information demand, social communication, as well as sharing and consumption of multimedia content. Because of this, many users turn to multiple of these online platforms, depending on their present individual needs, creating a heterogeneous multi-channel\footnote{Note that we choose to use the term \textit{channel} instead of \textit{stream} as we believe it to fit the archive-alike properties of multimedia collections such as YouTube, Flickr, and Wikipedia better.} environment within the social media landscape. 
While researchers have a general intuition about the nature of these channels (e.g. many current event detection systems are based on Twitter feeds \cite{osborne2012bieber, whiting2012hashtags}), studies of the exact distribution of information across multiple channels have not been in the scope of the multimedia community until now.
To the best of our knowledge, there does not exist a study on the behavior and lifecycle of trending topics across multiple media channels and on the correlations between topic categories and media channels over a one year observation period prior to this work. For example, the trending topic ``olympics 2012'' (the number one trend) manifests itself in multiple media channels which clearly exhibit different behavior (as illustrated in Figure \ref{fig:olympics}).

A second and much more challenging task is \textit{forecasting} the life cycle of a trending topic, i.e. forecasting the amount of user engagement towards it in the very moment they emerge.
These forecasts enable us to anticipate changing information needs of users ahead of time and to allocate limited resources in trend aware multimedia systems such as recommender systems \cite{diplaris2012socialsensor, roy2012socialtransfer} or video concept detection systems \cite{borth2012dynamic}). For example, such limited resources include time of content consumers and producers, screen estate and content recommendations on web pages \cite{roy2012socialtransfer}, processing time for training models in video concept detection \cite{borth2012dynamic}, and server load and replication capacity \cite{wang2012propagation}. 

While obviously of great use, forecasting trending topics is a very challenging problem since the corresponding time series usually exhibit highly irregular behavior also known as structural breaks that can lead to large forecasting errors and unreliable models \cite{clements2009forecasting}. The complexity of this task will be illustrated in more detail in Sec. \ref{sec:forecasting} (Figure \ref{fig:system_overview}).

We present a novel forecasting approach exploiting the assumption that semantically similar topics exhibit similar behavior in a nearest neighbor framework. Semantically similar topics are discovered by mining topics of similar type and category on DBPedia \cite{auer2007dbpedia}, a structured representation of the information on Wikipedia. We evaluate our fully automatic approach on a large-scale dataset of billions of views of several million articles on Wikipedia.

Summarizing, our contribution in this paper is three-fold. We present \textbf{(1)} the first comprehensive study of trending topics in three major social and online media channels with an observation period of one year, \textbf{(2)} a fully automatic forecasting technique for these trending topics based on a nearest neighbor approach exploiting semantic relationships between topics, and
\textbf{(3)} an empirical evaluation of our forecasting model over real-world user behavior on a large-scale Wikipedia dataset.

The rest of this paper is organized as follows. In Sec. 2, we outline work related to trending topics in multimedia and social media based forecasting. The multi-channel analysis of trending topics is presented in Sec. 3. Our novel forecasting technique is described in Sec. 4 before evaluating the approach in Sec. 5. Sec. 6 concludes this paper with a short summary and an outline for future work.

%% file: relatedwork.tex
\section{Related Work}
The overview over related work is divided into three parts. We first report work that analyzes or combines information from multiple media channels. Then, we focus on forecasting behavioral dynamics before finally presenting applications of trending topics to social multimedia systems.

\paragraph{Multi-Channel Analyses}
Detecting and tracking topics in news media has been studied since 1997 with the goal of understanding news in multiple languages and across multiple media channels (including television and radio sources) \cite{allan2002introduction}. 

Ratkiewicz et al.\ \cite{ratkiewicz2010traffic} argue that online popularity of specific topics (e.g. ``Barack Obama'') cannot in general be characterized by the behavior of individual news-driven events (e.g. ``Barack Obama inaugurated as U.S. President'') since the former might subsume many different news stories. They further find bursts in Wikipedia traffic to be correlated with bursts of Google search volume.

Wikipedia article views have also been correlated with behavior on Twitter to analyze the effectiveness of creating new content for breaking news (e.g. Japan earthquake) or updating pages at the moment of news (e.g. Oscar winners) \cite{whiting2012hashtags} and to filter spurious events on Twitter which indirectly resulted in the finding that Wikipedia lags behind Twitter by about two hours \cite{osborne2012bieber}.
Adar et al. correlate several behavioral datasets such as general queries from an observation period of one month from MSN and AOL search logs with the general idea of understanding past behavior to predict future behavior \cite{adar2007we}.

The work of Yang and Leskovec \cite{yang2011patterns} explores patterns of temporal variation in online media by clustering them to analyze the rise and fall of human attention with regard to certain phrases or memes. They report six distinct temporal patterns and find that weblogs trailed mainstream media by one or two hours for most of the considered phrases.

In contrast to the above, we explicitly focus our study on trending topics in online and social media channels over a long observation period. Similar to \cite{de2011find} using trending topics such as ``oil spill'' and ``iPhone'' to evaluate tweet selection, we utilize trending topics published by the media channels themselves and focus on multi-channel analysis and forecast.
We further analyze the correlation between topic categories and media channels and propose a fully automatic approach of forecasting trending topics in terms of time series (as opposed to visualization tools \cite{adar2007we} or predicting temporal clusters \cite{yang2011patterns}).

\paragraph{Forecasting Behavioral Dynamics}
Much research has been devoted to predicting economic variables such as auto sales, unemployment claims \cite{choi2012predicting} (``nowcasting''), or opening weekend box-office revenue for movies \cite{goel2010predicting}.
In the same way, popularity of online content has often been treated as a single variable (e.g. total number of views) instead of a time series. Either early popularity \cite{szabo2010predicting}, or content-based features such as publisher, subjectivity, or occurrence of named entities \cite{bandari2012pulse} are used to forecast eventual popularity. 
More recent work treats popularity of queries and clicked URLs of search engines as time series and uses state space models adapted from physics for forecasting \cite{radinsky2012modeling}.

The only work on forecasting Wikipedia article popularity known to the authors of this paper restricts itself to the case of featured articles on the main page and accounts for daily cycles in viewing behavior \cite{thij2012modeling}. However, trending topics lead to time series with unexpected shifts. These shifts are known as structural breaks in the field of econometrics and can lead to large forecasting errors and unreliable forecasting models \cite{clements2009forecasting}. 
A good example for such a structural break is Whitney Houston's death in February 2011 that caused 2000 times (!) more people to access her Wikipedia article than usual.
Clements and Hendry find that structural breaks (also described as parameter non-constancy) are the main cause of large, unexpected forecast errors in practice. 
Pooling or combining forecasts from different models has often been found to outperform individual forecasts and to increase forecast robustness. 

Forecasting trending topics is different in the sense that we require a fully automatic system and that often there is little historical information available (unlike for many economic variables of interest). For example, few people were aware of ``Costa Concordia'' before the ship sunk in January, 2012. Similarly, to make predictions about ``54th Grammy Awards'' one needs to understand the relationship of this event with previous ones such as previous instances of the Grammy Awards. 
We assume that semantically similar topics share characteristics and behavior and therefore could improve forecasting accuracy. This assumption has not yet been explored in previous work (e.g. \cite{radinsky2012modeling}). In addition, our proposed approach can forecast popularity of arbitrary articles that exhibit very diverse viewing dynamics. This differs dramatically from the experimental conditions in  \cite{thij2012modeling} which are characterized by the authors as ``nearly ideal''.

\paragraph{Multimedia Applications}
One application using social media in our target domain of social multimedia systems has been explored in \cite{jin2010wisdom}. This work uses the Flickr photo upload volume of specific topics to inform autoregressive nowcasting models for monthly political election results and product sales (i.e. the model requires the Flickr upload volume at time $t$ to produce a forecast for time $t$). Further, the Flickr queries relevant to the forecast subject of interest are chosen manually  (e.g. using ``Hillary'' instead of ``Clinton'' to avoid images by Bill Clinton for the 2008 Democratic Party presidential primaries). 

Another application that explicitly focuses on trending topics identified in different online and social media channels uses them to dynamically form and extend the concept vocabulary of video concept detection systems \cite{borth2012dynamic}. They further find trending topics to be strongly correlated with changes in upload volume on YouTube and demonstrate a visual classification of these trends.

SocialTransfer is another system that uses trending topics obtained from a stream of Twitter posts for social trend aware video recommendation \cite{roy2012socialtransfer}. Learning new associations between videos based on current trends is found to be important to improve the performance of  multimedia applications, such as video recommendation in terms of topical relevance and popularity. The popularity of videos is also used in \cite{wang2012propagation} to drive the allocation of replication capacity to serve social video contents. This work analyzes real-world video propagation patterns and finds that newly generated and shared videos are the ones that tend to attract the most attention (called temporal locality). They further formulate the challenge to estimate the videos' popularity for video service allocation for which they use the number of microblog posts that share or re-share the video. These insights are incorporated into the design of a propagation-based social-aware replication framework.
Two other research prototypes that seek to enhance the multimedia consumption experience by extracting trending topics and events from user behavior on the Web are SocialSensor \cite{diplaris2012socialsensor} and TrendMiner \cite{samangooei2012trendminer}. The first one emphasizes the real-time aspects of multimedia indexing and search over multiple social networks for the purpose of social recommendations and retrieval. TrendMiner focuses on real-time methods for cross-lingual mining and summarization of large-scale stream media and use cases in financial decision support and political and monitoring.

%% file: analysis.tex
\section{Multi-Channel Analysis}
\label{sec:analysis}
This section introduces the dataset used for the analysis of trending topics across media channels and presents an analysis of their temporal characteristics as well as insights into the relationship between channels and topic categories.

\subsection{Trending Topics Dataset} \label{subsec:trending_topic_dataset}
To facilitate a comprehensive analysis of trending topics across multiple channels, we crawl the top trends from three major online media channels on a daily basis. With the intent to capture people's communication needs, search patterns, and information demand we choose to retrieve trends from Twitter, Google, and Wikipedia, respectively. We further cluster similar trends across time \textit{and} channels to later be able to compare the different manifestations of a particular trend across multiple channels. 

\begin{table}[t]
\centering
\scriptsize
\begin{tabular}{rlrl}
\hline\hline\noalign{\smallskip}
 & \textbf{Topic} & & \textbf{Topic}\\ 
\noalign{\smallskip}\hline\noalign{\smallskip}
1. & olympics 2012 & 11. & christmas\\
2. & champions league & 12. & steve jobs\\ 
3. & iphone 5 & 13. & manhattan\\ 
4. & whitney houston & 14. & academy awards\\ 
5. & mega millions numbers & 15. & formula 1\\ 
6. & closer kate middleton & 16. & justin bieber\\ 
7. & facebook & 17. & joe paterno died\\
8. & costa concordia & 18. & battlefield 3\\ 
9. & black friday deals & 19. & muammar gaddafi dead\\ 
10. & superbowl & 20. & ufc\\ 
\noalign{\smallskip}\hline\hline
\end{tabular}
\caption{International Top 20 Trending Topics during Sep.\ 2011 -- Sep.\ 2012. Please note that the US presidential election was in Nov. 2012.}\label{tab:top_trends}
\end{table}
\normalsize

\paragraph{1. Raw Trending Topic Sources}
As motivated above, we use Google, Twitter and Wikipedia as a starting point by retrieving a ranked list of popular terms from 10 different sources on a daily basis: five Google channels (Search and News for USA and Germany as well as the Trends feed), three Twitter channels (daily trends for USA and Germany as well as the Daily Trends feed), and two Wikipedia channels (popular articles in the English or German language). For each of these feeds we retrieve 10-20 ranked topics (110 topics per day). In total the dataset covers the observation period Sep.\ 2011 -- Sep.\ 2012 and contains roughly 40,000 potentially overlapping topics. 

\paragraph{2. Unification and Clustering of Trends}
To connect multiple instances of the same trending topic across time and media channels we perform a unification and clustering of the individual topics obtained in the previous step. First, we map the individual topic strings to a corresponding Wikipedia URI by selecting the top-most Wikipedia result when performing a Google search for that topic (this approach was found to be more robust than more direct methods on Wikipedia \cite{borth2012dynamic}). Then we cluster two URIs together if their Levenshtein or edit distance is below a certain threshold (set to 0.35 $\times$ word length). This process can be thought of as a greedy version of a hierarchical agglomerative clustering using the Levenshtein distance. The method allows us to unify topics such as ``super bowl time'', ``super bowl 2012'', and ``superbowl'' into a single cluster. Overall this results in 2986 clusters or individual trends. A cluster is now represented by its most prominent member and will be referred to as a \textit{trending topic} for the rest of this work.


\paragraph{3. Ranking Trending Topics}
To reason about the popularity of these trending topics we assign them trend scores based on the following method: For each each day and for each of our 10 feeds, we record the rank at which a topic appears. These ranks are combined using Borda count, obtaining a score for each day that is assigned to the topic's cluster (from the previous step). To measure the impact of a trending topic (cluster) over its overall lifetime, we define its global trend score by the sum of its daily scores over the observation period. The top trends with respect to this global trend score are shown in Table \ref{tab:top_trends}.

\begin{table}
\centering
\begin{tabular}{lrr}
\hline\hline\noalign{\smallskip}
 & \textbf{\#Topics} & \textbf{\#Sequences}\\
\noalign{\smallskip}\hline

\textbf{Total} & 200 & 516\\
\hline
\textbf{Google} & 191 (95.5\,\%) & 445 (86.2\,\%)\\
\textbf{Twitter} & 118 (59.0\,\%) & 232 (45.0\,\%)\\
\textbf{Wikipedia} & 69 (34.5\,\%) & 108 (20.9\,\%) \\
\hline
\textbf{G \& T} & 115 (57.5\,\%) & 174 (33.7\,\%)\\ 
\textbf{G \& W} & 66 (33.0\,\%) & 86 (16.7\,\%)\\
\textbf{T \& W} & 43 (21.5\,\%) & 57 (11.0\,\%)\\
\hline
\textbf{G \& T \& W} & 42 (21.0\,\%) & 48 (9.3\,\%)\\
\noalign{\smallskip}\hline\hline
\end{tabular}
\caption{Number of analyzed trend topics and sequences for Google (G), Twitter (T), and Wikipedia (W).}
\label{tab:channel_summary}
\end{table}

\subsection{Lifetime Analysis}
Some trending topics occur multiple times within our one year observation period. For example sport topics such as ``Champions League'' are discussed whenever important games are played.
To allow for an analysis of when channels begin and stop featuring particular topics, we divide a trending topic into multiple sequences. For the Champions League example, we obtain one sequence per round of the tournament. Formally, topics are broken up into sequences such that their daily trend scores are non-zero for at least two out of any three adjacent dates, i.e. we compensate for ``score gaps'' of at most one day but form multiple sequences for gaps of two or more days. We perform an analysis of the lifetime of trending topics in the different channels for the top 200 trends (based on their global trend score) that were split into 516 sequences based on the described procedure. Table \ref{tab:channel_summary} summarizes the resulting number of topics and sequences for the different channels and their combinations. Note that we map the ten individual observed channels to their respective sources: Google, Twitter, and Wikipedia. The Google channel has the largest coverage of the trending topic sequences in our dataset (86.2\,\%). About 9.3\,\% of trending topic sequences occur in all three channels and between 11.0\,\% and 33.7\,\% occur in two channels.

\begin{figure}
\centering
\subfloat[Lifetime of all channels.]{\includegraphics[width=0.45\columnwidth]{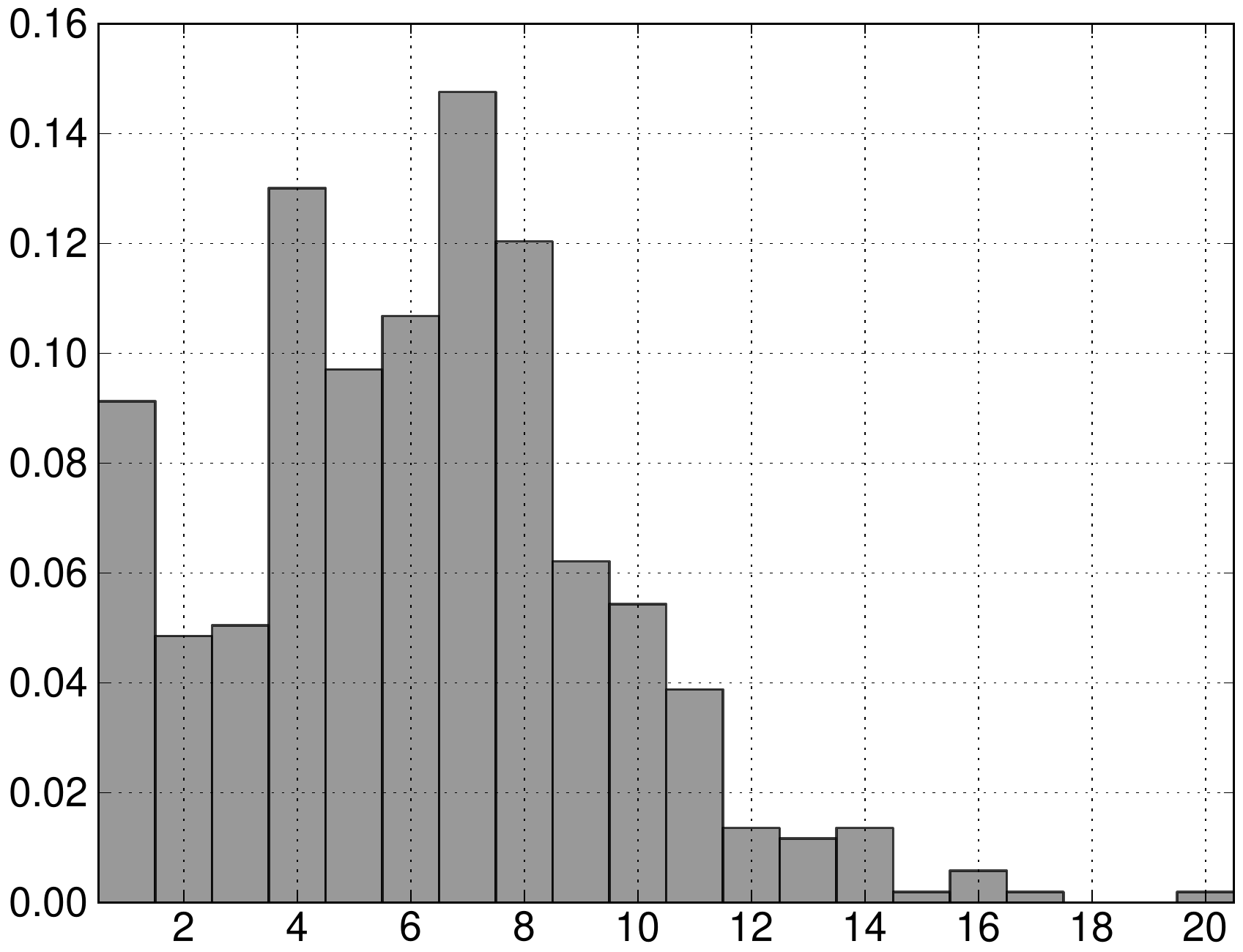}}\hfill
\subfloat[Lifetime for Google]{\includegraphics[width=0.45\columnwidth]{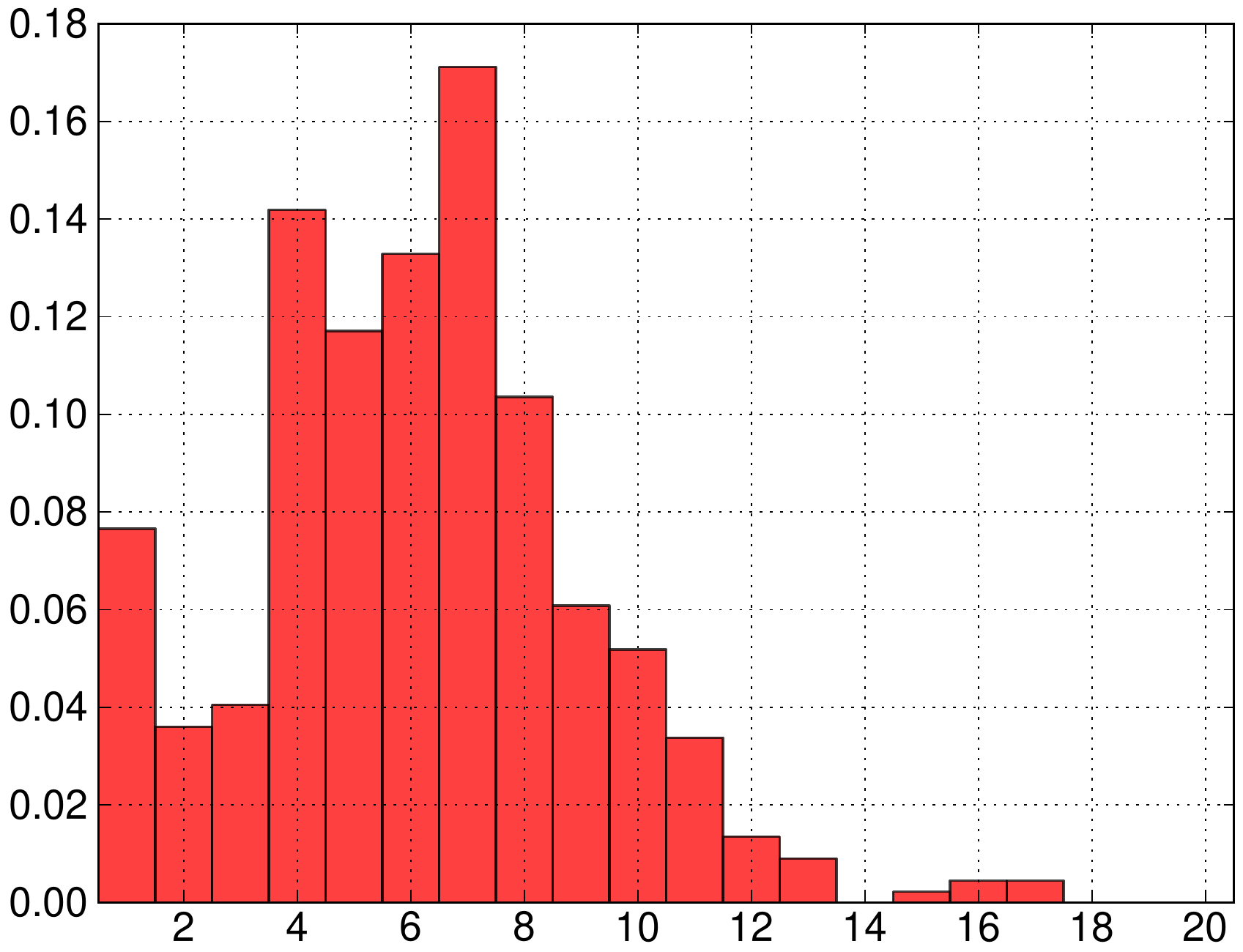}}\\
\subfloat[Lifetime for Twitter]{\includegraphics[width=0.45\columnwidth]{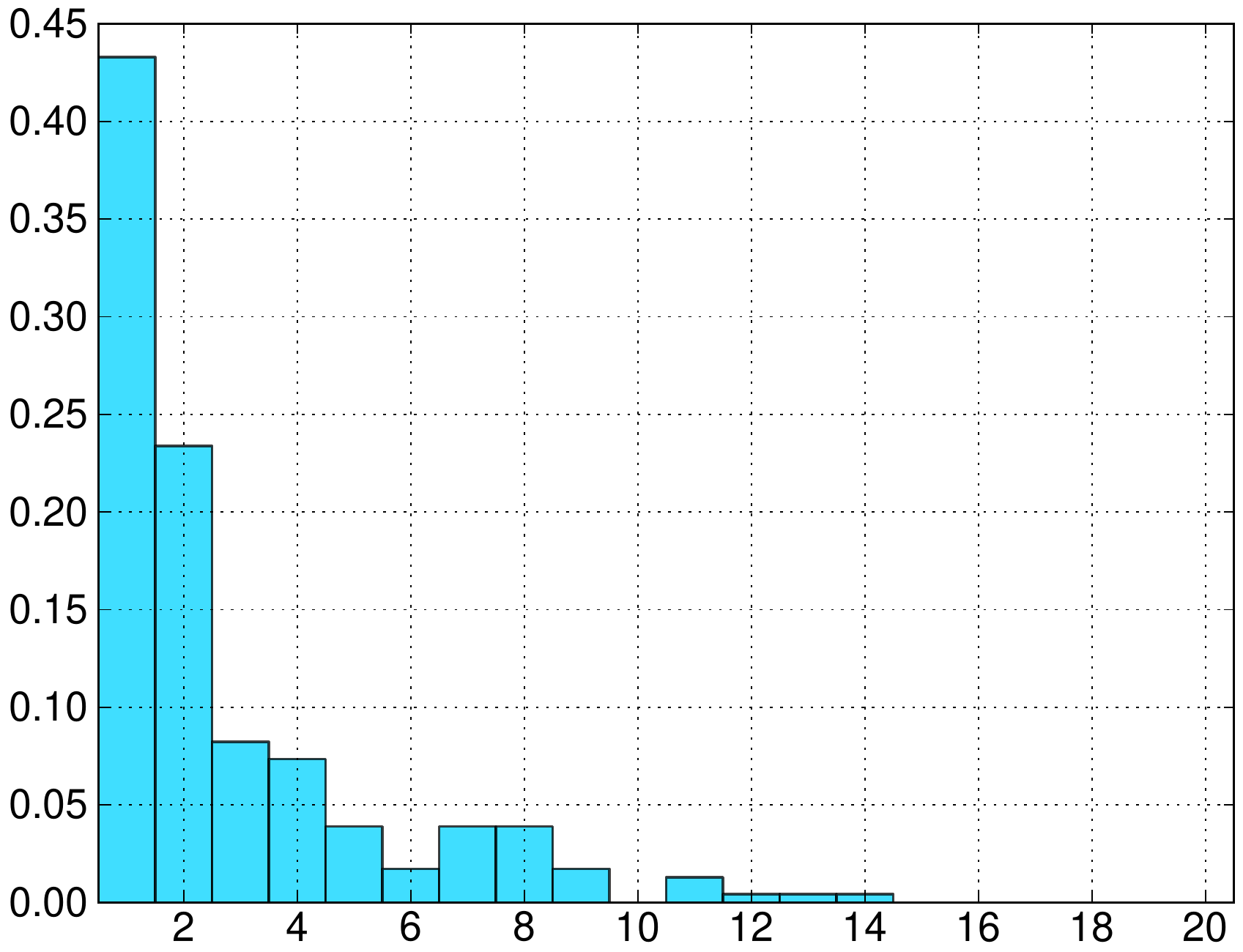}}\hfill
\subfloat[Lifetime for Wikipedia]{\includegraphics[width=0.45\columnwidth]{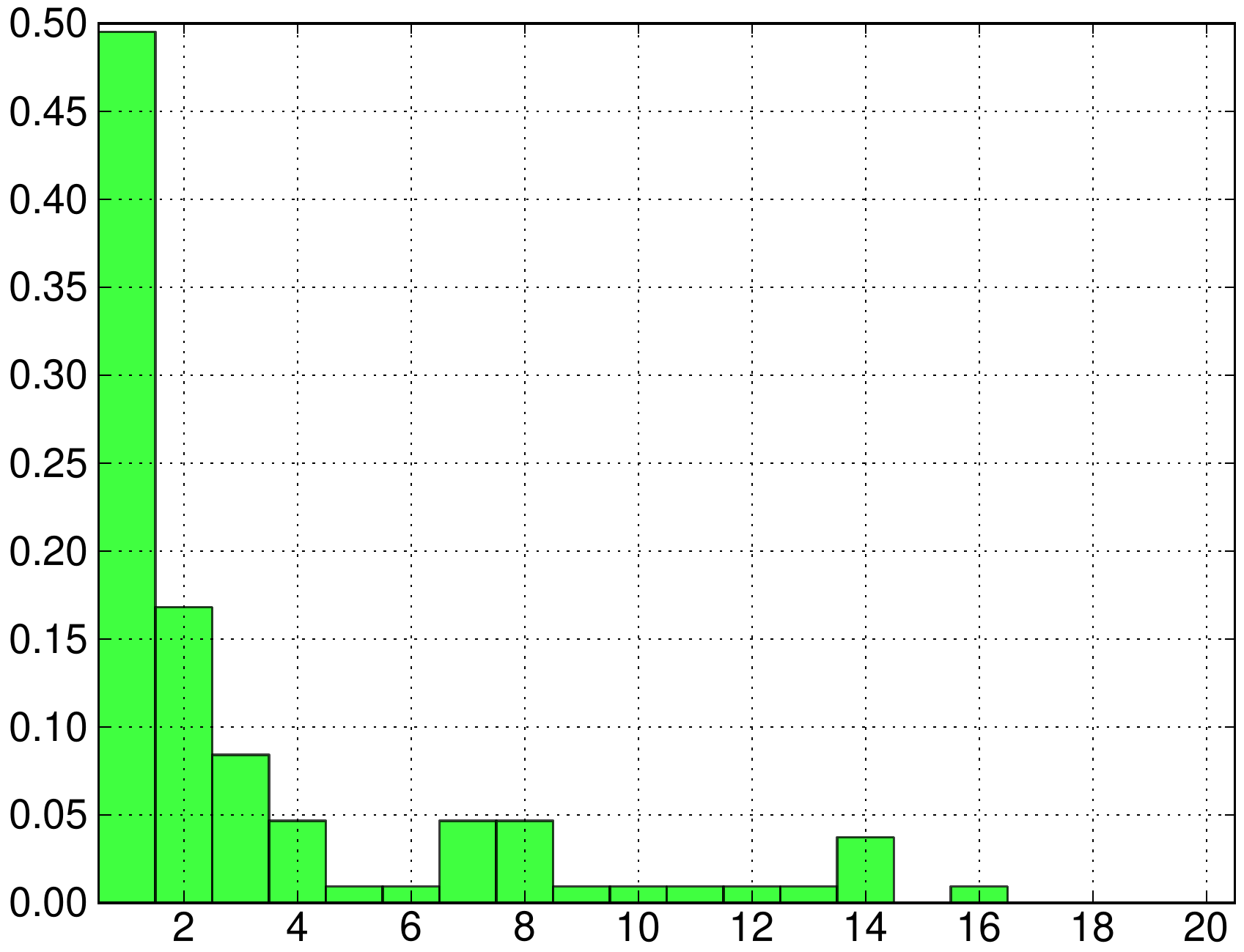}}\\
\caption{Lifetime in days for top 200 trending topics.}
\label{fig:lifetime_histograms}
\end{figure}

We first deal with the question how long trending topics survive in general and whether there are differences in lifetime in the different media channels. In our case, lifetime is defined as the number of consecutive days with positive trend scores. Histograms of the lifetime of trending topics are shown in Figure \ref{fig:lifetime_histograms}. We observe that the trending topics in our dataset rarely survive for more than fourteen days with most trending topics having a lifetime of less than nine days. 
Since Google covers a large share of top trends, the distribution for the channel looks very similar to the overall distribution. The lifetime of topics on Twitter is much shorter complying with our expectations of the ephemerality of trends in this channel -- about two thirds of the top trends only survive for one or two days. Interestingly, the distribution looks similar for Wikipedia. This raises the question when and why exactly people are turning to Wikipedia to satisfy their information needs.

For trending topics that occur in at least two of the three channels we record the day on which the trend starts, the day on which the trend peaks (defined by the trend score defined earlier), and the day on which the trend ends again. We then define the \textit{delay} between two channels as the difference between the start/peak/end dates in channel X and the start/peak/end days in channel Y. Note that a positive delay means that the first channel X is slower, i.e. trends tend to start later in channel X than in channel Y. The mean delays for start, peak, and end are summarized in Table \ref{tab:lifetime_delay}. Interestingly, there are only marginal differences in starting delays (first of the three numbers) between the three channels with Twitter and Wikipedia being slightly faster than Google. These results are not very surprising considering that Osborne et al. found Twitter to be around two hours faster than Wikipedia \cite{osborne2012bieber} -- a difference almost impossible to observe in data of daily granularity. A much stronger effect is observed for peak and end delays (second and third number). Both Twitter and Wikipedia tend to peak more than two days before Google. The picture is even clearer when looking at the end delays where Twitter and Wikipedia lead Google by three and four days respectively. Overall, these results add to the ephemerality of the Twitter and Wikipedia. However, through this analysis we found that some of the Google channels seem to be intentionally delayed or averaged, i.e. still displaying the top stories after several days.
Since Google Trends is known not to be artificially delayed we added comparisons with this channel to Twitter and Wikipedia to our analysis. This channel peaks around the same time as Twitter and Wikipedia and trends tend to end around half a day after Twitter and more than one day after Wikipedia.
\begin{table}
\centering
\small
\begin{tabular}{ccc} 
\hline\hline\noalign{\smallskip}
& \textbf{Twitter} & \textbf{Wikipedia}\\
\hline
\textbf{Google} & -0.20 / 2.23 / 3.64 & -0.41 / 2.35 / 4.16 \\ 
\hline
\textbf{Twitter} & -- & 0.07 / 0.36 / 0.21\\ 
\hline
\textbf{Google Trends}& 0.09 / 0.04 / 0.41 & -0.42 / -0.32 / 1.25\\ 
\noalign{\smallskip}\hline\hline
\end{tabular}
\caption{Mean delay in days between pairs of media channels (start/peak/end). Positive delay means that the ``row channel'' is slower than the ``column channel''.}
\label{tab:lifetime_delay}
\end{table}

\subsection{Topic Category Analysis across Channels}
\begin{figure}
\centering
\includegraphics[width=\columnwidth]{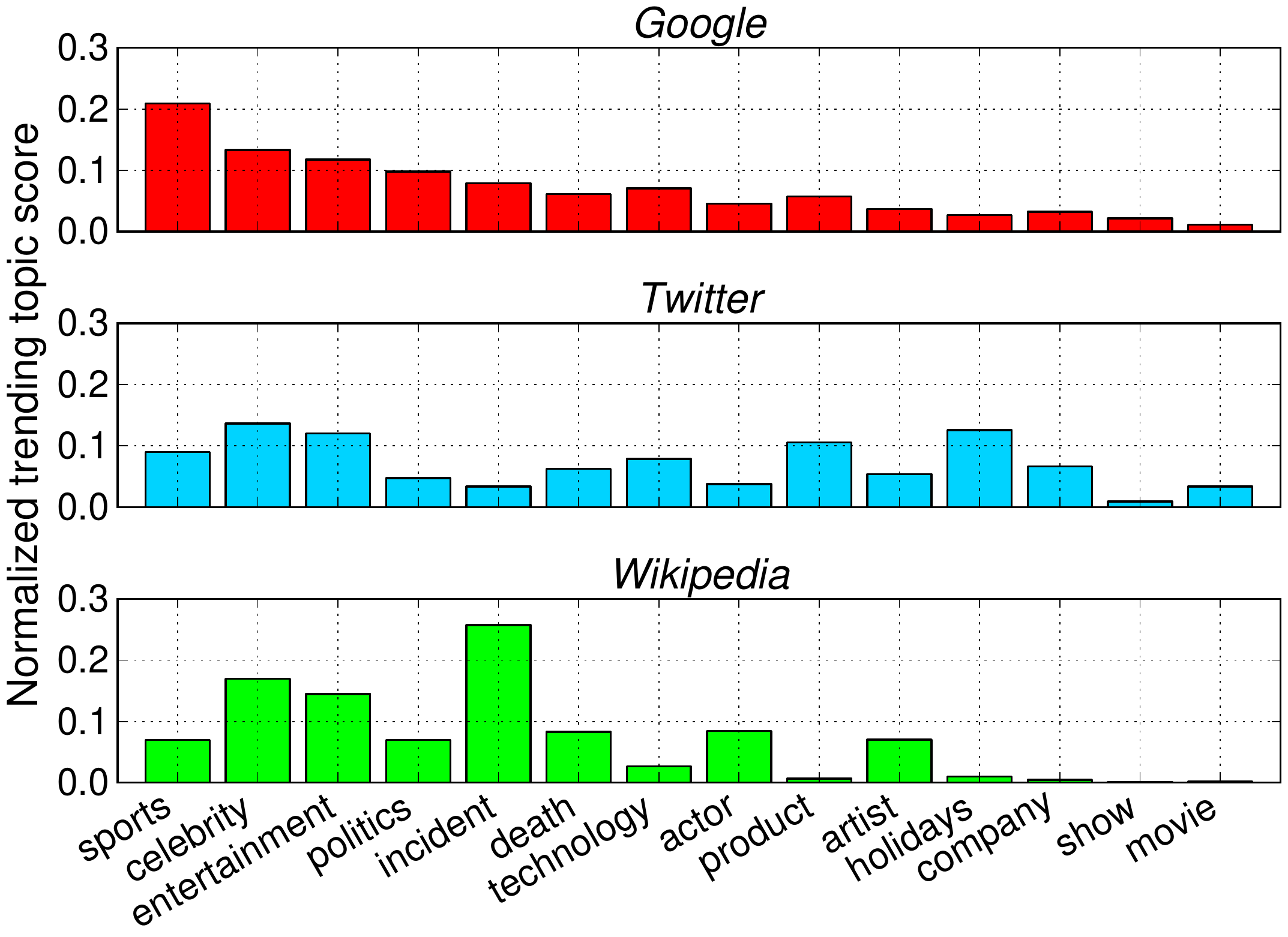}
\caption{Normalized distribution of trending topic scores over trend categories in the individual channels. Note that channels tend to specialize in certain categories, e.g. Wikipedia for incidents.}\label{fig:channel_category_correlation}
\end{figure}

Considering trend aware multimedia applications it is critical to understand why people to turn to specific channels. To shed light on what kind of topics are the most popular in the individual channels we manually annotate the 200 top trends with categories. The categories are chosen by examining the main themes of trends we found in our dataset. Please refer to Table \ref{tab:trend_categories} for more information about the individual categories, their descriptions, and examples. Note that a trending topic might be assigned to multiple categories, i.e. the death of a popular music artist would be assigned to celebrity, entertainment, death, and artist to make an extreme example (similar to \cite{adar2007we}). The engagement with respect to the different categories in a media channel is measured as follows: For each trend within the channel we assign its score to all of its categories. Finally, we normalize the scores for each channel, e.g. to account for the dominance of Google for the scores overall. The resulting distribution over categories is displayed in Figure \ref{fig:channel_category_correlation}. 

\begin{figure}[t]
\centering
\includegraphics[width=\columnwidth]{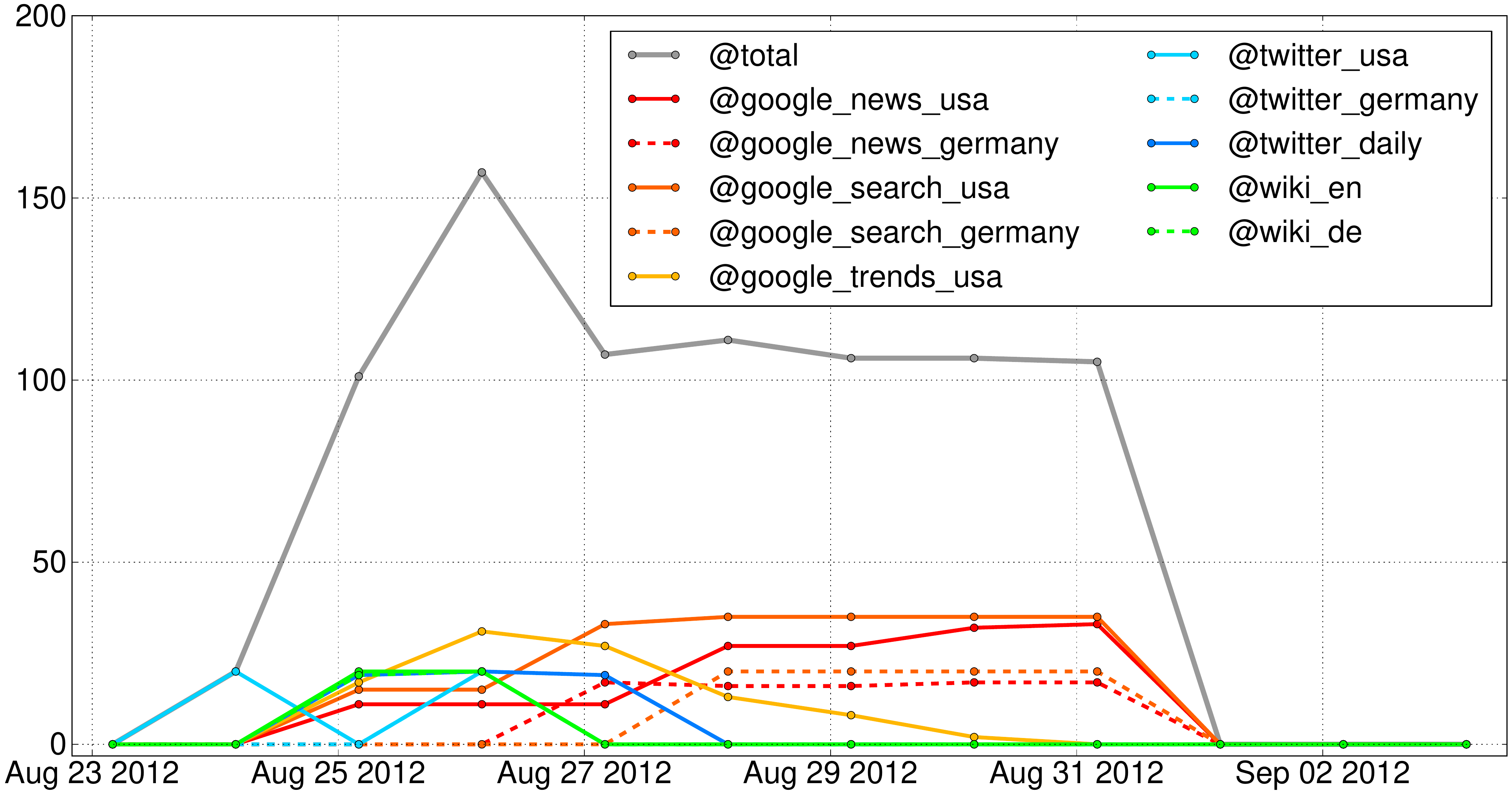}
\caption{Neil Armstrong received much attention when he died on August 25, 2012. Note that the peak on Wikipedia (green) is at the beginning of the trending topic period whereas news coverage continued for a couple more days.}\label{fig:neil_armstrong}
\end{figure}

We can observe that the channels tend to specialize in certain topic categories.
For example, the most popular category in Google is sports.
A large share of the scores further is assigned to celebrity and entertainment categories. Google also has the highest relative share (15\,\%) for politics. Twitter also features many trends in the celebrity and entertainment categories. Interestingly, it has the highest relative shares of trends related to products, companies and technology. One reason might be that many Twitter users are technology affine early adopters that like to share their thoughts on new products. Another interesting finding is that over 20\,\% of the scores on Twitter are assigned to the holidays category. We hypothesize that holiday related trends are big on Twitter as many people tag their posts and pictures with the same hashtags such as \#christmas or \#thanksgiving. Wikipedia clearly shows a specialization for categories that involve people and incidents such as disasters or the death of celebrities. Contrary to the intuition that Wikipedia is a slowly evolving channel which people use to read up on complicated topics, especially when also considering the temporal properties of the Wikipedia channel from the analysis above, we believe that many users use Wikipedia during these trends and events to learn about or remind themselves about related topics. For example, when Neil Armstrong died on August 25, 2012 many people viewed the corresponding Wikipedia article e.g. to learn about who he was (American astronaut), what he was famous for (first person on the moon), and how old he was when he died (82) (see Figure \ref{fig:neil_armstrong}). Note that this trending topic also mistakenly subsumes the lifetime ban from sports competitions of Lance Armstrong on August 24, 2012.

\begin{table}[t]
\centering
\scriptsize
\begin{tabular}{l>{\raggedright}p{1.5cm}r>{\raggedright\arraybackslash}p{2.3cm}}
\hline\hline\noalign{\smallskip}
 \textbf{Category} & \textbf{Description} & \textbf{\#Seq} & \textbf{Examples}\\
\hline\noalign{\smallskip} 
 \textbf{sports} & sports events, clubs, athletes & 52 & olympics 2012, champions league, bayern muenchen\\ 
 \textbf{celebrity} & person with prominent profile  & 49 & steve jobs, kim kardashian,  michael jackson, neil armstrong\\ 
 \textbf{entertainment} & entertainers, movies, TV shows & 39 & grammys, emmys, heidi klum\\ 
 \textbf{politics} & politicans, parties, political events, movements & 32 & paul ryan, occupy, christian wulff\\ 
 \textbf{incident} & an individual occurrence or event & 27 & costa concordia, hurricane isaac, virginia tech shooting\\  
 \textbf{death} & death of a celebrity & 22 & whitney houston, joe paterno died, neil armstrong\\ 
 \textbf{technology} & product or event related to technology & 20 &  iphone 5, ces, nasa curiosity\\ 
 \textbf{actor} & actor in TV show or movie & 18 & lindsay lohan, michael clarke duncan, bill cosby\\ 
 \textbf{product} & product or product release & 15 & ipad, windows 8, diablo 3\\ 
 \textbf{artist} & music artist & 15 & justin bieber, miley cyrus, beyonce baby\\ 
 \textbf{holidays} &  day(s) of special significance & 11 & halloween, thanksgiving, valentines day\\  
 \textbf{company} & commercial business & 10 & apple, chick fil a, megaupload\\  
 \textbf{show} & TV show & 7 & x factor, wetten dass, the voice\\  
 \textbf{movie} & a motion picture & 6 & dark knight rises, hunger games, the avengers\\  
 \noalign{\smallskip}\hline\hline
\end{tabular}
\caption{List of categories of trending topics. Note that a trending topic might be assigned to multiple categories.}
\label{tab:trend_categories}
\end{table}
\normalsize

%% file: forecasting.tex
\section{Forecasting of Trending Topics}
\label{sec:forecasting}


As motivated in the introduction, anticipating changing information needs and shifts in user attention towards emerging trends is critical for trend aware multimedia systems. However, forecasting trending topics is a very challenging problem since the corresponding time series usually exhibit highly irregular behavior (structural breaks) when the topic becomes ``trending''.
In the following, we will be using the multi-channel pipeline described in Sec. \ref{subsec:trending_topic_dataset} as a robust trigger for trending topics but will perform time series forecasting utilizing historic time series data (see Sec. \ref{subsec:wikipedia_dataset} for more details).
In this section, we propose a novel forecasting approach that combines time series from multiple semantically similar topics.

\begin{figure*}[t]
\begin{center}
\includegraphics[width=\textwidth]{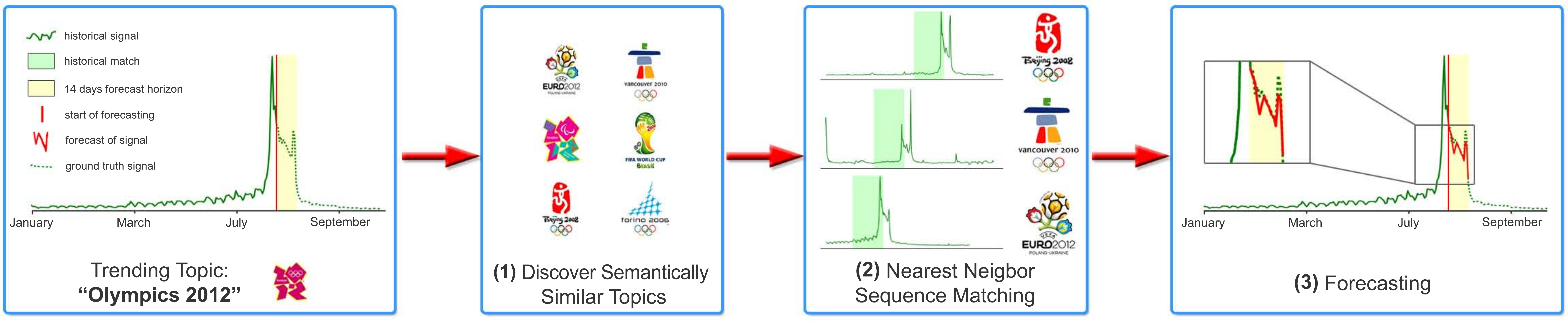}
\end{center}
\caption{System overview of our proposed forecasting approach. For a given trending topic we first discover semantically similar topics which we search for patterns of similar behavior which are then used to produce a forecast.}\label{fig:system_overview}
\end{figure*}

A conceptual overview of our system is presented in Figure \ref{fig:system_overview}. On the very left you can see the trending topic for the Summer Olympics 2012 along with its Wikipedia page view statistics in 2012. The task to be solved is to forecast the number of page views for a period of 14 days (yellow area) from the day indicated by the red line. This forecast start is triggered by the emergence of a corresponding trending topic in our observed channels (see Sec.\ \ref{sec:analysis}).  Note that the time series exhibits complex behavior such as the smaller peak at the end of the forecasting period which most likely corresponds to the closing ceremony event on this day.
Based on the finding that semantically similar events can exhibit very similar behavior, the first step is to automatically discover related topics such as previous Summer and Winter Olympics or FIFA/UEFA soccer championships (as illustrated by the second box). 
The second step in Figure \ref{fig:system_overview} shows patterns of user engagement for Summer Olympics 2008, the Winter Olympics 2010, and the UEFA Euro championship 2012 that were found to match the historic behavior of the 2012 Olympics best.
These patterns show certain commonalities such as a second peak for closing ceremonies or final matches.
The identified sequences from the previous step are then combined to a forecast shown at the very right.
In the following, these individual steps are explained in more detail.

Please note that we are neither able to nor do we attempt to predict incidents such as natural disasters or sudden deaths of celebrities in advance. However, even for unpredictable events like these, the patterns of user attention once this event has happened can be forecasted e.g. by taking previous instances of natural disasters or celebrity deaths into account.

%

\subsection{Discovering Semantically Similar Topics} \label{subsec:discovering_semantically_similar_topics}
The basis for our forecasting approach is the hypothesis that semantically similar topics exhibit similar behavior. 
One example supporting this hypothesis is given in Figure \ref{fig:system_overview} which clearly shows similar patterns of user engagement during the different sport events. For a given trending topic, we discover semantically related topics using DBPedia \cite{auer2007dbpedia}, a dataset containing structured information about several million named entities extracted from the Wikipedia project. 
Given category information (via \texttt{dcterms:subject}) and type (via \texttt{rdf:type}) from DBPedia, we build a set of (up to 200) semantically similar topics that share the most categories or types with a given trending topic. For example, the Wikipedia URI for ``Olympics 2012'' 
is assigned to the categories \texttt{Sports\_\-festivals\_\-in\_\-London}, \texttt{Scheduled\_\-sports\_\-events}, \texttt{2012\_\-Summer\_\-Olympics}, and \texttt{2012\_\-in\_\-London}. Its types include \texttt{Event}, \texttt{SportsEvent}, \texttt{Olympics}, and \texttt{OlympicGames}, e.g. the 2012 Summer Paralympics share four categories with the Olympics 2012.

Formally, we use a topic set $\mathcal{T}_{sim}$ which includes all discovered similar topics. For later comparisons, we also define a topic set $\mathcal{T}_{self}$ that only includes the trending topic itself and $\mathcal{T}_{gen}$ which contains a wide variety of general topics (in our case the top 200 trending topics). In the following, we will use $\mathcal{T}$ as a placeholder for one of these topic sets. The elements will be referred to as similar topics or $\text{sim}_j \in \mathcal{T}$.

\subsection{Nearest Neighbor Sequence Matching}
\label{subsec:nnseqmatching}
Obviously not all time series corresponding to the discovered similar topics look the same. Therefore, we search within these time series for sequences that match historical behavior of our trending topic. For example, historical topics such as the 1896 Summer Olympics have gained very limited attention over the last years and are therefore unlikely to be representative for the large amount of engagement towards the 2012 Summer Olympics. To pick the right instances to inform our forecast, we compare a short history window of the trending topic to be forecasted, i.e. the viewing statistics for the last two months, to all partial sequences of the same length of similar topics in our topic set $\mathcal{T}$.

To capture this step in formal terms, let $S_{\text{topic}}[t] \in \mathbb{R}$ be the time series for the given topic at time $t \in \mathbb{N}$. Further define $S_{\text{topic}}^{t_0}[t] := S_{\text{topic}}[t_0 + t]$ as the shifted version of the time series (used for aligning multiple series below). In the following, we assume that we want to forecast $S_{\text{topic}}$ with a horizon of $h$ days starting at time $t_0 \in \mathbb{N}$.

Given a topic set $\mathcal{T}$ from the previous step, we define a sequence candidate set $C(t_0)$ that includes all possible shifted time series $S_{\text{sim}_j}^{t}$: $C(t_0) = \{S_{\text{sim}_j}^{t} |\, \forall \text{sim}_j \in \mathcal{T}, \;\forall t:\; t \leq t_0-h\}$. The condition  $t \leq t_0-h$ ensures that we never use information more recent than $t_0-h$, to allow for a forecast of $h$ days, i.e. we do not utilize future information.

Given this candidate set, we now search for the $k$ members $S_{\text{sim}_i}^{t_i}$ ($i=1,\dots,k$) that are the best matches for our time series of interest ($S_{\text{topic}}^{t_0}$). Note that these nearest neighbors are already correctly aligned through shifting the time series $S_{\text{sim}_i}$ by a corresponding $t_i$. Note that the same similar topic $\text{sim}_i$ can occur multiple times (e.g. for repetitive signals).

Formally, the nearest neighbor set becomes $N^k(S_{\text{topic}}^{t_0}) = \{ S_{\text{sim}_1}^{t_1}, \dots, S_{\text{sim}_k}^{t_k} \}$ where the $S_{\text{sim}_i}^{t_i}$ are the $k$ distinct elements that are smallest wrt.\ $d(S_{\text{topic}}^{t_0}, S_{\text{sim}_i}^{t_i})$ for all $S_{\text{sim}_i}^{t_i} \in C({t_0})$. Here, $d(\cdot, \cdot)$ is a distance metric between both time series which, in our case, only depends on a short history window of the time series.
An interesting question is whether the metric should be scale invariant and in which form and to what degree. In Sec.\ \ref{sec:evaluation}, we compare three different distance metrics: (1) (squared) \texttt{euclidean} distance: ($d(x, y) = \sum_{i=1}^n (x_i - y_i)^2$), (2) euclidean distance on normalized sequences $x_i' = (x_i - \mu)/\sigma$ (where $\mu, \sigma$ are mean and standard deviation estimated from the respective time series) referred to as \texttt{musigma}, and (3) a fully scale invariant metric (called \texttt{y\_invariant}) that was proposed in \cite{yang2011patterns} ($\min_\gamma d(x,\gamma \cdot y)$).

\subsection{Forecasting}
Even the best matching sequences identified in the previous step might not be a perfect fit for the time series to be forecasted. Therefore, we rescale the matching sequences such that they all match on the last observed value of $S_{\text{topic}}^{t_0}$. This ensures that the forecast will be a continuous extension of past behavior. Now, the forecast $\mathcal{F}$ becomes the median over scaled versions of the sequences from the previous step ($N^k(S_{\text{topic}}^{t_0})$).
We represent the days that we want to forecast by $\tau \in [0,\dots,h-1]$ where again $h$ is the forecasting horizon (usually $h=14$).
Our forecast can then be formally described as $$\mathcal{F}(S_{\text{topic}}^{t_0})[\tau] = \median_{S_{\text{topic}'}^{t'} \in N^k(S_{\text{topic}}^t)} (\alpha(S_{\text{topic}}^{{t_0}}, S_{\text{topic}'}^{t'}) \cdot S_{\text{topic}'}^{t'}[\tau])$$ where $\alpha(S_{\text{topic}}^{{t_0}}, S_{\text{topic}'}^{t'}) = S_{\text{topic}}^{{t_0}}[-1](S_{\text{topic}'}^{t'}[-1])^{-1}$ adjusts the scale of nearest neighbor time series based on the last observed score. 
In practice, we limit $\alpha$ to a limited interval (e.g. $[0.33, 3.0]$) for robustness.
We further evaluate our approach using the average instead of the median for forecasts. However, as presented in Sec.\ \ref{sec:evaluation} the median proves to be more robust.

%

%% file: evaluation.tex
\section{Evaluation}
\label{sec:evaluation}
In this section, first the Wikipedia page views dataset is described before presenting quantitative results for the forecasting approach proposed in the previous section.

\subsection{Dataset Description} \label{subsec:wikipedia_dataset}
We evaluate our proposed forecasting approach on a large-scale dataset of page views on Wikipedia, an online collaborative encyclopedia that has become a mainstream information resource worldwide and is frequently used in academia \cite{ratkiewicz2010traffic, whiting2012hashtags, osborne2012bieber, thij2012modeling}.
Reasons for this particular choice of social media channel were (a) the public availability of historical views data necessary to build forecasting models (hourly view statistics for the last five years), (b) the size of the dataset allowing a comprehensive analysis of our proposed method across a wide range of topics (over 5 million articles), and (c) previous results show that user behavior on Wikipedia (bursts in popularity of Wikipedia pages) is well correlated with external news events \cite{ratkiewicz2010traffic} and therefore support our proposed method's ability to generalize to different social media channels. Although we only provide results on Wikipedia our approach could be applied to any online and social media channel for which historic data is available.

\begin{table}[h]
\centering
\scriptsize
\begin{tabular}{>{\raggedright}p{2.5cm}>{\raggedright\arraybackslash}p{5.0cm}}
\hline\hline\noalign{\smallskip}
\textbf{Trending Topic} & \textbf{Nearest Neighbor Topics}\\
\noalign{\smallskip}\hline 
\textbf{2012 Summer Olympics} & \textbf{2008 Summer Olympics, UEFA Euro 2012, 2010 Winter Olympics} :: 2016 Summer Olympics, 2014 FIFA World Cup, 2006 Winter Olympics\\ 
\textbf{Whitney Houston} & \textbf{Ciara, Shakira, Celine Dion, Brittany Murphy, Ozzy Osbourne} :: Alicia Keys, Paul McCartney, Janet Jackson\\ 
\textbf{Steve Jobs} & \textbf{Mark Zuckerberg, Rupert Murdoch, Steve Jobs} :: Steve Wozniak, Bill Gates, Oprah Winfrey\\ 
\textbf{Super Bowl XLVI} & \textbf{Super Bowl, Super Bowl XLV, Super Bowl XLIV} :: Super Bowl XLIII, 2012 Pro Bowl, UFC 119\\ 
\textbf{84th Academy Awards} & \textbf{83rd Academy Awards, 82nd Academy Awards} :: List of Academy Awards ceremonies, 81st Academy Awards \\ 
\textbf{Battlefield 3} & \textbf{Mortal Kombat, FIFA 10, Call of Duty: Modern Warfare 2, Portal, Duke Nukem Forever} :: Call of Duty: Modern Warfare 3, Call of Duty 4: Modern Warfare, Pro Evolution Soccer 2011\\ 
\textbf{Joe Paterno} & \textbf{Terry Bradshaw, Joe Paterno, Jack Ruby, Paul Newman, Jerry Sandusky} :: Lane Kiffin, Donna Summer, Joe DiMaggio\\ 
\textbf{54th Grammy Awards} & \textbf{53rd Grammy Awards, 52nd Grammy Awards, 54th Grammy Awards} :: 51st Grammy Awards, 2012 Billboard Music Awards, 2012 MTV Europe Music Awards\\
\noalign{\smallskip}\hline\hline
\end{tabular}
\caption{Selected trending topics along with their nearest neighbor topics using category and type information on DBPedia (step 1). The ones chosen by nearest neighbor sequence matching (step 2) are in bold print. In some cases the topic itself can be used for forecasting, e.g. if the time series contains repetitive patterns.}
\label{tab:similar_topics}
\vspace{-1em}
\end{table}

The raw Wikipedia viewing statistics are published\footnote{\url{http://dumps.wikimedia.org/other/pagecounts-raw/}} by the Wikimedia foundation. We obtained hourly view statistics starting on January 1, 2008 (2.8 TB compressed in total). We aggregate these logs to daily viewing statistics where we drop URIs that have been viewed less then 25 times on that day. We believe that this does not introduce any bias since trending topics tend to accumulate volumes several orders of magnitude higher. For each day this results in approx. 2.5 million URIs attracting 870 million daily views. In total, the English and German Wikipedia features more than five million articles that can be used for forecasting. During data preparation we noticed that Wikipedia did not record view counts for a few short periods of time during the last five years for which we interpolate linearly.

Note that we use this dataset for forecasting of historical time series data (actual Wikipedia viewing statistics) while the multi-channel pipeline described in Sec. \ref{subsec:trending_topic_dataset} serves as a robust trigger for trending topics.

\subsection{Experiments}
We structure our experiments along the three main building blocks of our proposed approach to compare design choices for the individual methods independently (recall Fig. \ref{fig:system_overview}).

\paragraph{1. Discovering Semantically Similar Topics}
We evaluate the influence of discovering semantically similar topics in two ways. First, we present qualitative results by showing retrieved similar topics for a few trending topics. Second, we evaluate this part indirectly by comparing forecast performance (i.e. through the third step) of using semantically similar topics from DBPedia to using a general set of topics (the top 200 trending topics themselves). 

Subsets of semantically similar topics for the top trends are shown in Table
\ref{tab:similar_topics} (as generated by the approach described in Sec. \ref{subsec:discovering_semantically_similar_topics}). Note that we successfully identify similar events or previous instances for events like the Olympics, the Super Bowl, UFC events, or the Grammy Awards. We are also able to discover similar people like similar music artists, entrepreneurs, athletes, and even people that died from the same cause (such as lung cancer for Joe Paterno, Jack Ruby, and Paul Newman). On average, we retrieve 95 semantically similar topics per trending topic.

\paragraph{2. Nearest Neighbor Sequence Matching}
The main design choice when matching sequences from similar topics to a short history window of our time series is the choice of the distance metric. As introduced in Sec.\ \ref{subsec:nnseqmatching} we compare \texttt{euclidean}, \texttt{musigma}, and \texttt{y\_invariant}. 

Our experimental setup for this part as well as the forecasting looks as follows. We use the trending topics acquired in Sec.\ \ref{sec:analysis} as a trigger for forecasting, i.e. for each of the 516 sequences of the top 200 trending topics (see Table \ref{tab:channel_summary}), we make a prediction for a horizon of 14 days, starting on the day they first emerge. We chose this window of 14 days since this represents the maximum lifetime for most trending topics (see Figure \ref{fig:lifetime_histograms} (a)). Our method is evaluated by beginning with a forecast of 14 days, then a 13 day forecast after one day and so on (X axis in error plot).

To compare the distance metrics, we measure the quality of the (first) nearest neighbor returned by this metric by its similarity to the actual viewing statistics over the next 14 days (similar to the forecast setting but directly using the nearest neighbor as the forecast). The quality of this match is captured by an error metric. In this work, we use root mean squared error defined as $RMSE = \sqrt{ \frac{1}{n} \sum_{t=1}^n (A_t - F_t)^2}$ where $A_t$ is the actual value and $F_t$ is the forecast value. As a relative error metric we choose mean absolute percentage error (MAPE) defined as $MAPE = \frac{100\,\%}{n} \sum_{t=1}^n |\frac{A_t - F_t}{A_t}|$. Note that unlike \cite{radinsky2012modeling} we measure error on the actual view counts instead of normalizing by the total views for each day as we believe this glosses over large relative errors (i.e. even larger trends might only account for $10^{-5}$ of the daily views yielding very small error rates for virtually any forecast).

\begin{figure}
\centering
\includegraphics[width=\columnwidth]{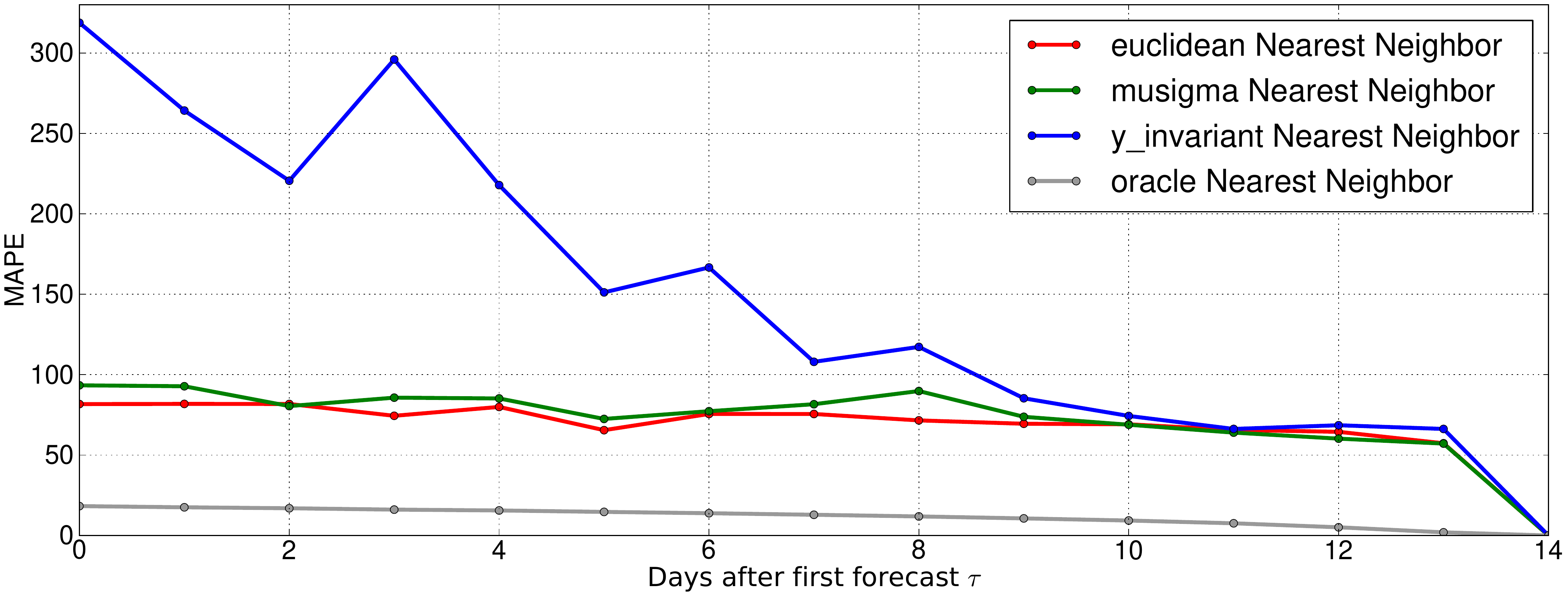}
\caption{A comparison of Nearest Neighbor Distance Metrics. Euclidean distance performs at least as good as its more complex normalized counterparts.}\label{fig:nn_matching_comparison}
\end{figure}

The results are depicted in Figure \ref{fig:nn_matching_comparison} and show the average MAPE error (lower is better) between the nearest neighbor and the actual viewing statistics (ground truth). We also show oracle performance by picking the best match for the ground truth from all sequence candidates. First, note that oracle performance is well above 0\,\% error as there might be no perfect match among the sequence candidates. Further, we observe that \texttt{y\_invariant} can retrieve poor quality matches whereas the simple \texttt{euclidean} distance performs equally or better than the other two metrics. Apparently, the invariance of \texttt{musigma} and \texttt{y\_invariant} does not help in this task. Therefore, we use the \texttt{euclidean} metric for all following experiments. Further note that the best matches have between 82 and 319\,\% error illustrating the high complexity of the task.

\paragraph{3. Forecast}
As introduced above we measure forecasting performance by forecasting the next 14 days for each of the 516 sequences of the top 200 trending topics at the point in time when they first emerge.
Our approach is compared to several baselines that use a short history window of the time series itself (similar to \cite{radinsky2012modeling}): a \textit{naive} forecast (tomorrow's behavior is the same as today's) and a \textit{linear trend} based on the last 14 days. We further compare to the \textit{average trend} and \textit{median trend} in our trending topics dataset as a baseline that includes multiple time series. Note that this average and median trend are computed from $\mu/\sigma$-normalized time series since the average/median of actual view counts are actually very far from most trending topics. To still be able to compute the error for actual view count prediction we then de-normalize the \textit{average trend} and \textit{median trend} baselines with the parameters of the time series to be predicted.
Note that while we performed the following experiments for different numbers of neighbors (for the nearest-neighbor-based methods) we only report the results for $k=3$ which performed best by a small margin.

\begin{table}
\centering
\small
\begin{tabular}{llccc}
\hline\hline\noalign{\smallskip}
& \multirow{2}{*}{\textbf{Method}} & \multicolumn{3}{c}{\textbf{RMSEs in 1000}}\\
& & $\tau = 0$ & $\tau = 3$ & $\tau = 7$ \\ \hline 
\multirow{4}{*}{\textbf{Baselines}} & naive &
	63.2 & 33.1 & 17.4 \\
& linear trend &
	86.9 & 48.5 & 23.1 \\
& average trend &
	49.3 & 25.9 & 19.9 \\
& median trend &
	48.1 & 24.9 & 18.1 \\
\hline\hline
\multirow{4}{*}{\textit{Self}} & average &
	46.0 & 23.7 & 18.0 \\
& average\_scaled &
	44.6 & 21.9 & 15.5 \\
& median &
	46.1 & 23.8 & 17.7 \\
& median\_scaled &
	44.9 & 22.3 & 15.5 \\
\hline
\multirow{4}{*}{\textit{Gen}} & average &
	45.7 & 22.9 & 16.1 \\
& average\_scaled &
	45.7 & 22.5 & 14.1 \\
& median &
	41.4 & 21.2 & 15.4 \\
& median\_scaled &
	40.1 & 19.5 & 12.8 \\
\hline
\multirow{4}{*}{\textit{Sim}} & average &
	41.4 & 18.8 & 14.0 \\
& average\_scaled &
	39.6 & 17.1 & 11.6 \\
& median &
	42.1 & 19.9 & 14.0 \\
& median\_scaled &
	41.0 & 17.9 & 11.5 \\
\hline\hline

\end{tabular}
\caption{RMSE forecasting error for our baselines as well as methods using only the trending topics itself (\textit{Self}), a general set of topics (\textit{Gen}), or similar topics (\textit{Sim}).}
\label{tab:rmse_forecasting_error}
\end{table}

The RMSE forecasting errors are summarized in Table \ref{tab:rmse_forecasting_error} which reports the results for a number of instances of our proposed forecasting approach. \textit{Self}, \textit{Gen}, and \textit{Sim} refer to the different topic sets  $\mathcal{T}_{self}$, $\mathcal{T}_{gen}$, and $\mathcal{T}_{sim}$ from which the nearest neighbor sequences are chosen (as described in the last section). On average, the RMSE of our best method is about 9-48k views closer to the actual viewing statistics than the different baseline methods.
Further, we observe that taking the median tends to perform about as good or better than taking the average, using scaled nearest neighbor is better than unscaled neighbors, and that using semantically similar topics (\textit{Sim}) is better than using a general set of topics (\textit{Gen}) which in turn is better than restricting oneself to a single time series (\textit{Self}).

\begin{figure}
\centering
\includegraphics[width=\columnwidth]{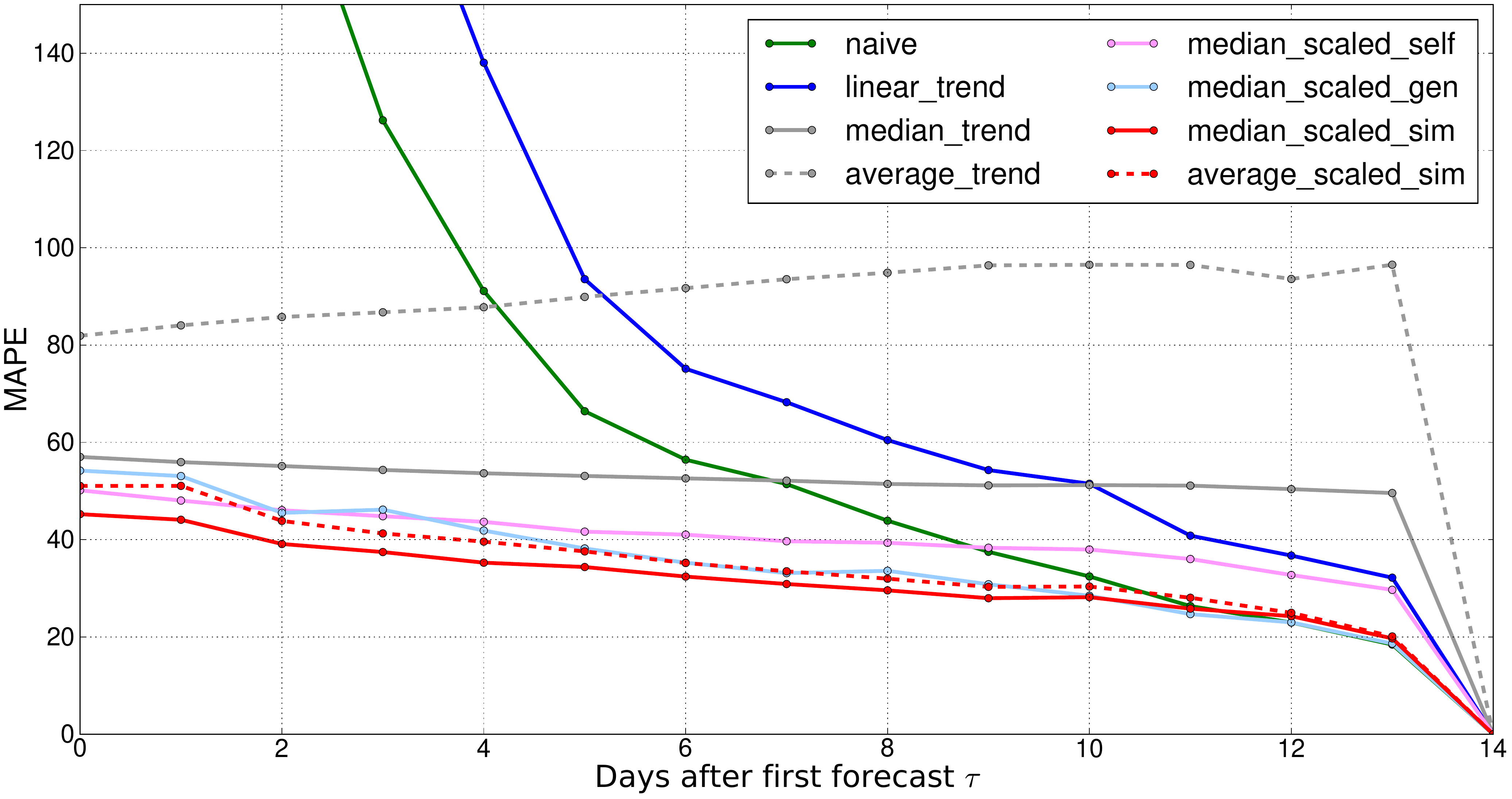}
\caption{MAPE forecast error moving through a 14 day window, e.g. the error at 4 depicts the forecasting error of the following 10 days.}\label{fig:mape_errors}
\end{figure}

However, RMSE error has the disadvantage that it is dominated by the most popular trending topics with the largest view counts. Therefore, we choose the MAPE measure for the remaining analysis as this relative error metric is comparable across trending topics. Additionally, because MAPE errors can become disproportionally large (e.g.\ forecasting 1000 views when it is actually only 100 results in a 900\,\% relative error), we drop obvious outliers (5\,\%) and report the average error for the remaining sequences. We only show the results for the baselines and the best performing methods from Table \ref{tab:rmse_forecasting_error} to be able to visually distinguish them in the error plot.
Again, the methods are evaluated by beginning with a forecast of 14 days, then a 13 day forecast after one day etc. as illustrated by the X axis in Figure \ref{fig:mape_errors}. 
We can observe that the proposed approach including median and scaling clearly outperforms all baselines as well as other instances of our framework. Our proposed method achieves a mean average percentage error (MAPE) of 45-19\,\%, a relative improvement over our baselines of 20-90\,\%.

We further performed a per-category analysis of the forecasting results (using the categories from Table \ref{tab:trend_categories}). Summarizing, the forecasting performance across categories is similar. Product, company, artist and technology categories perform slightly better, and death, actor, sports, holidays and politics perform slightly worse. For all categories, the forecast quality improves with time ($\tau$).

\begin{figure}
\centering
\includegraphics[width=\columnwidth]{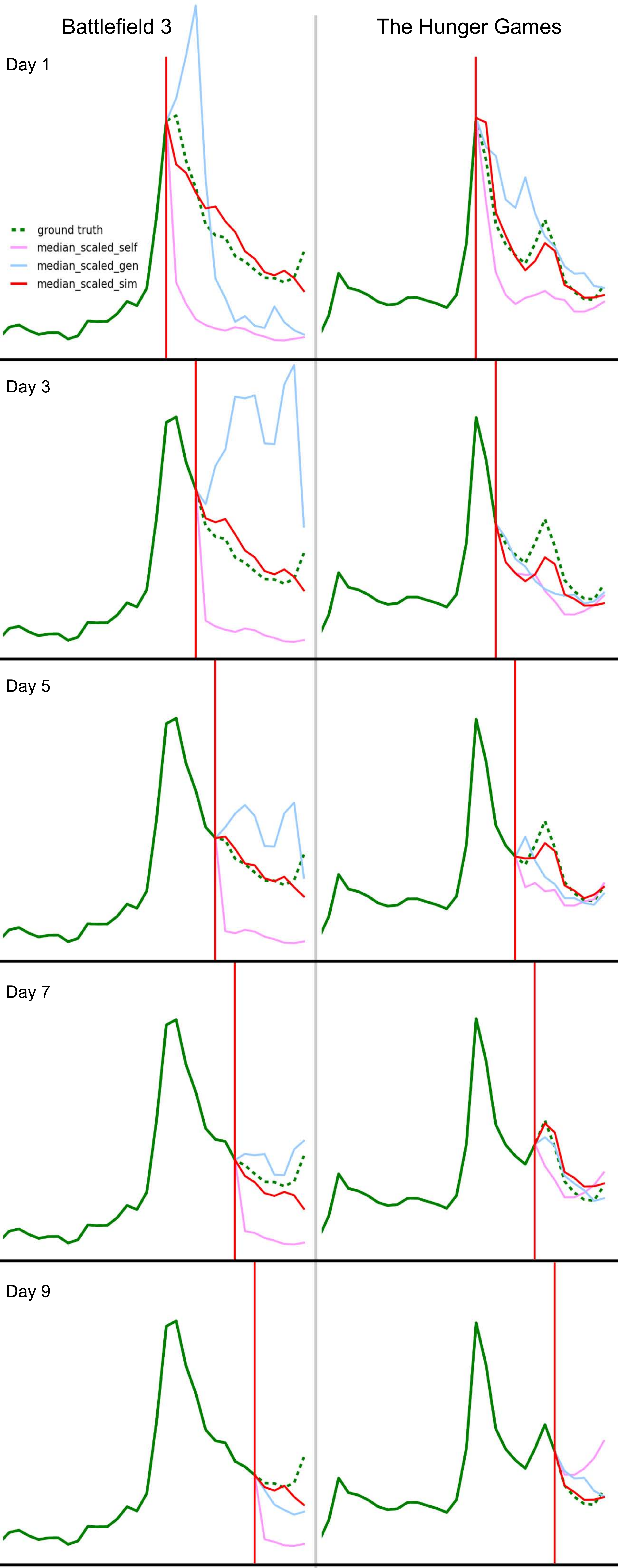}
\caption{Visualization of trending topic forecasts for ``Battlefield 3'' and ``The Hunger Games''. Each column depicts multiple forecasts at different points in time (as indicated by the vertical red line).}\label{fig:qualitative_results}
\end{figure}

Example forecasts for two trending topics, ``Battlefield 3'' (computer game) and ``The Hunger Games'' (novel and movie), are given in Figure \ref{fig:qualitative_results}. Each column depicts multiple forecasts at different points in time (as indicated by the vertical red line). Our proposed method \texttt{median\_scaled\_sim} (\textit{Sim}) is compared to its variants only using the trending topic itself (\textit{Self}) or using a set of general topics (\textit{Gen}). Note that in many cases, only using information from a single time series is not enough. Further, using semantically similar topics (\textit{Sim}) leads to more accurate forecasts that e.g. are able to capture multiple peaks (such as in the Hunger Games example). 

%% file: conclusion.tex
\section{Conclusion}
We present a comprehensive study of trending topics in three major social and online media streams with an observation period of one year. Our findings include that trends on Twitter and Wikipedia are more ephemeral than on Google, both rising and declining rapidly for newly emerging topics. Further, we find that the observed media channels tend to specialize in specific topic categories. The second key contribution is a fully automatic forecasting technique of trending topics exploiting semantic similarity between topics. Lastly, we evaluate our approach on a large-scale dataset of Wikipedia viewing statistics and show empirically that forecasts by the proposed approach are about 9-48k views closer to the actual viewing statistics than baseline methods and achieve a mean average percentage error of 45-19\,\% for time periods of up to 14 days, a relative improvement over baselines of 20-90\,\%.


As future work, the proposed model could be extended to explicitly detect and exploit seasonality as well as incorporate global changes in viewing statistics.
In a more general ranking framework, the nearest neighbor sequence matching part could further be extended by e.g.\ the number of common semantic types and categories. 

